\documentstyle [psfig]{article}
\newcommand{\de}{\delta}
\newcommand{\De}{\Delta}

\newcommand{\ga}{\gamma}

\newcommand{\La}{\Lambda}
\newcommand{\la}{\lambda}
\newcommand{\Om}{\Omega}

\newcommand{\be}{\begin{equation}}
\newcommand{\ee}{\end{equation}}
\newcommand{\gsim}{\stackrel{>}{\sim}}

\newcommand{\bea}{\begin{eqnarray}}
\newcommand{\eea}{\end{eqnarray}}
\newcommand{\bean}{\begin{eqnarray*}}
\newcommand{\eean}{\end{eqnarray*}}

\newcommand{\cd}{\cdot}

\title{CMB anisotropies caused by gravitational waves:\\
        A parameter study}
\author{Ruth Durrer$^1$ and Tina Kahniashvili$^2$}
\begin{document}
\maketitle
$^1$Universit\'e de Gen\`eve, D\'epartement de Physique Th\'eorique,
4, quai E. Ansermet, CH-1211 Gen\`eve 4, Switzerland\\
$^2$Department of Astrophysics, Abastumani Astrophysical Observatory,
Kazbegi ave. 2a,  380060 Tbilisi, Georgia

\begin{abstract}
\vskip 10pt
 Anisotropies in the cosmic microwave background radiation due to
gravity waves are investigated. An initial
spectrum of gravity waves may have been induced during  an
epoch of inflation. We study the  propagation of such a spectrum
in different Friedmann backgrounds with hot and cold dark matter, radiation
 and, possibly, a cosmological constant. We  calculate its
imprint as anisotropies on the cosmic microwave background.

We also take into account that massless particles can
source gravity waves by their anisotropic stresses. We consider
general mixed dark matter models with and without cosmological
constant. For a given, scale invariant input spectrum of gravity
waves, we determine
the dependence of the resulting spectrum of
CMB anisotropies on the different parameters of the model.

\end{abstract}
{\bf Keywords: } Cosmology: cosmic microwave background; Gravitational
waves; Cosmology: large-scale structure of the Universe.

\vspace{2cm}
\section{Introduction}
The theoretical and observational determination of anisotropies in the
cosmic microwave background radiation (CMB) has recently attracted a lot
of attention. One has justified hopes to measure the CMB anisotropies
to a precision of a few
percent or better within the next ten years. Furthermore,
 if initial fluctuations are induced during a
primordial inflationary period and no external sources induce
perturbations at later times, CMB anisotropies can be calculated by
linear cosmological perturbation theory
 to very good accuracy. Since the
detailed results depend not only on the primordial spectrum but also
on the parameters of the cosmological model considered, the 
anisotropy spectrum may provide a mean to determine these parameters
to an accuracy of a few percent.

This problem has been extensively investigated mainly for scalar
perturbations \cite{Berkelygang}.

As it is well known,  also tensor perturbations can by generated 
during inflation. They play an  important role: 
As shown in \cite{maa96},
the presence of gravitational waves can crucially change the
theoretical predictions of cluster abundance, which is an important test
of cosmological models. In particular, power spectra of mixed dark
matter (MDM) models
normalized to the  COBE 4-year data \cite{bun97}
 provide cluster abundances higher than observed.
This is one of the difficulties of standard MDM models.
Taking into account a gravitational wave contribution, 
this inconsistency can be circumvented. Hence, the question how 
gravitational wave contributions
 depend on model parameters is very important.

Here we discuss the model dependence of
anisotropies due to gravitational waves for models with a total
density parameter $\Omega=1$ which are thus spatially flat. However, we
vary the contributions of cold dark
matter (CDM), hot dark matter (HDM) and a cosmological constant, which
are given in terms of the parameters
$\Omega_C~,~\Omega_H$ and $\Omega_{\Lambda}$.
We also vary the number of degrees of freedom for massless neutrinos
and hot particles.

We consider a fixed input spectral index $n=0$ from inflation. For a
general input spectrum $\langle|h(t_{in},k)|^2\rangle =A(k)^2k^{-3}$, 
our output spectrum
$\langle|\dot {h}(t,k)|^2\rangle$  has to be multiplied by $|A(k)|^2$.

In Section~2 we present the perturbation equations and describe the
models considered in this work. In Section~3 we discuss our results
and we conclude in Section~4. The non-trivial relation between the
temperature anisotropy spectrum, $C_\ell$, and the metric fluctuation
spectrum  $\langle h_{ij}(t,{\bf k})h_{lm}(t',{\bf k})^*\rangle$ for
tensor perturbations is derived in the appendix. \vspace{0.3cm}\\
{\bf Notation:}\hspace{0.3cm}The Friedmann metric is given by
$a^2(-dt^2+\ga_{ij}dx^idx^j)$, where $a$  denotes the scale factor, $t$
is conformal time, and $\gamma$ is the metric of a three space with
constant curvature $K$. We shall consider a spatially flat universe,
the case $K=0$ exclusively.
 An over-dot stands for derivative with
respect to conformal time $t$, while prime denotes the derivative with
respect to $kt\equiv x$.
\vspace{1cm}

\section{The models}
\vspace{0.5cm}
The basics of linear perturbations of Friedmann universes are
 discussed in \cite{Review}. We shall adopt the notation of
\cite{Review} in this work.
We want to determine the evolution of tensor perturbations  in spatially
 flat Universes which contain a fraction of cold dark matter (CDM), hot
dark matter (HDM) and a cosmological constant $\La$, such that:
$\Om_0= \Om_H+\Omega_C+\Omega_{\Lambda}=1$, where $\Om_{\bullet}$ denotes
the density parameter today, i.e. at $t_0$. We neglect the
contribution of photons, massless neutrini and baryons (which may be
included in CDM) to $\Om_0$.

         The expansion of the Universe is described by the Friedmann
equation for the scale factor:
\be
\left({ \dot{a} \over a}\right)^2={8\pi \over 3}G{\rho}_M a^2+{1 \over 3} \Lambda a^2
\label{exp}
\ee
where $\rho_M$ is the total density of matter in the Universe,
$\rho_M=\rho_C+\rho_H+\rho_{\nu_0}+\rho_{\gamma}$. Here
$\rho_{\gamma}$ denotes the
density of photons,  $\rho_{\nu_0}$ is the density of massless
neutrini and $\rho_C$, $\rho_H$ denote the densities of CDM and HDM
respectively.

The metric element for a Friedmann universe with  tensor
perturbations  is given by:

\be
ds^{2}=a^{2}(t)(-dt^2+(\delta_{ij}+2h^T_{ij})
dx^idx^j)     \label{pert}
\ee
where we choose  $c=1$, $t$ is conformal time and
$a(t)$ denotes the scale factor. For tensor perturbations, the metric
fluctuations $h^T_{ij}$
satisfy the conditions
\be
h^{Ti}_i=0,~~~~~h^{Tj}_{i}k^{i}=0
\label{cond} \ee
where ${\bf k}$ is the wave vector which may be set  equal to $(0,0,k)$,
such that the conditions (\ref{cond}) reduce to
\be
h^T_{11}=-h^T_{22},~~~~~\mbox{and}~~~~~~ h^T_{i3}=h^T_{3i}=0~.
\ee

We describe dynamics of tensor perturbations  in a medium containing
collision-less particles, whose anisotropic stresses are not damped
by collisions. As long as  the collision-less component is
relativistic, it provides a source for
gravitational waves. The evolution equation for tensor perturbations
of the  metric is given by \cite{Review}:
\be
{\ddot{h}^T}_{ij}+2{\dot a \over a}\dot{h}^T_{ij}+
 k^2 h^T_{ij}=8\pi G a^2p  \Pi_{ij} ~.
\label{hprime}
\ee
Here $p$ is the pressure of the collision-less component and
$\Pi$ denotes the tensor contribution to the anisotropic
stresses, which in our case are due to the presence of relativistic,
collision-less particles.
Denoting the tensor part of the perturbed
distribution function of the collision-less component by $F$, $\Pi$ is given
by
\be
\Pi_{ij} = {1 \over pa^4} \int { {v^4 \over q}
(n_{i}n_{j}-
1/3 \delta_{ij})F dv d{\Omega}}       \label{Pi}
\ee
where $n_{i}$ is a spatial unit vector, denoting the photon
directions, $v$ is the redshift corrected velocity and $q$ the
redshift corrected
energy of the collision-less particles (see \cite{Review}).
In the case of massless particles (massless neutrini)
$q \equiv v$.
Liouville's equation leads to the following perturbation equation for
$F$ \cite{Review},
\be
q \dot{F}+v n^jk_jF=qv n^i
n_j \dot{h}^{Tj}_i{df \over dv}~,
\label{Liou} \ee
 $f$ denotes the  unperturbed distribution function.

The set of Eqs.~(\ref{exp}) to (\ref{Liou}) fully
describes the evolution of tensor
perturbations  in media containing perfect fluids and collision-less
particles. In our models we have in principle three kinds of
collision-less particles: Hot dark matter, massless neutrini and,
after recombination, the photons.
Studying the initial conditions, we
shall find that for the growing mode anisotropic stresses are extremely
small on super horizon scales. However, when the scales relevant
for tensor CMB anisotropies ($\la\gg t_{dec}$) enter the horizon,
$t\gg t_{dec}$ HDM particles are already non relativistic. We may
thus neglect their contribution to anisotropic stresses.
Hence, we just consider the pressure anisotropy from massless neutrini
and, after recombination, from the photons themselves.

For massless particles we can  simplify Eqs.~(\ref{Pi}) and
(\ref{Liou}) by introducing the brightness perturbation  $M$
\be
M \equiv {{4\pi} \over {\rho_{\nu_0}a^4}} \int_0^{\infty} {Fv^3dv}
\label{M}
\ee
In terms of $M$ Liouville's equation (\ref{Liou}) becomes:
\be
\dot{M} + in^{j}k_{j}M=
-4n^{l}n^{j}{\dot{h}^T}_{l j}
\label{Liou.rel}
\ee
and the anisotropic stresses are given by:
\be
\Pi_{i j}={3 \over {4\pi}} \int {(n_{i}n_{j}-
1/3 \delta_{{i}{j}})Md\Omega}~.
\label{Pi.rel}
\ee
Equations~(\ref{hprime}),(\ref{Liou.rel}),(\ref{Pi.rel}) for
perturbations and eqn.~(\ref{exp}) for scale
factor together with the usual equations determining
$\rho_C,~\rho_H~\rho_{\nu_0}$ and $\rho_\gamma$ form the closed system
of ordinary differential equation which we have solved.

We assume standard inflation according to which the  initial amplitude of
gravitational waves is independent of scale {\it i.e.,}
$\langle|h(t_{in},k)|^2\rangle \propto k^{-3}$.

Each solution of Eqs.~(\ref{hprime}),(\ref{Liou.rel}) and
(\ref{Pi.rel}) can be presented as a sum of
growing and decaying modes and an infinite number of  modes
corresponding to perturbations of the collision-less medium
\cite{JETF}. We are only interested in the growing mode which is given
by the initial condition
$h^T_{i j}(t \rightarrow 0)=$const. and
 $\dot{h}_{i j}(t \rightarrow 0)=0$.

If $\dot{h}_{ij}=0$, Eq.~(\ref{Liou.rel}) does not admit a tensor
contribution to $M$. In this case,  $M\propto \exp(i{\bf n}\cdot{\bf k})$
and all components of the induced anisotropic stress normal to ${\bf k}$
vanish. Therefore, the correct (tensorial) initial condition for $M$ is
$M(t \rightarrow 0)=0$ and also  $\Pi_{ij}(t \rightarrow 0)=0$.

These initial values remain unchanged as long as $kt\ll 1$.
Assuming, e.g. spherical polarization and a flat spectrum from inflation,
at some early time,
$kt\ll 1$ for all wavelengths considered, we thus choose the initial
conditions
\be
\langle|h^T_{11}|^2\rangle = \langle|h^T_{12}|^2\rangle =
\langle|h|^2\rangle = A^2k^{-3},~~~~\Pi_{11} = \Pi_{12} = 0,~~~~M=0 ~,
\label{init}
\ee
where $A$ is the amplitude of gravitational waves.
It is easy see that on superhorizon scales ($kt \ll 1$),
$h=$const. and the evolution
of gravitational waves and as a result $\Delta T/T$ are
independent of the model parameters. For scales $kt \approx 1$,
the metric perturbations begin to oscillate and eventually ($kt\gg
1)$ damp away (see Figs.~1 and 2). The non-zero $\dot{h}$ then induces anisotropic
stresses via Eqs.(\ref{Liou.rel}) and (\ref{Pi.rel}). Very often, these
anisotropic stresses have been neglected in the literature. Here we
find that their effect is indeed very small. There is typically
about 1\% additional
damping due to the loss of some gravitational wave energy into
anisotropic stresses.

The main model dependence is the modification of the damping term
$(\dot{a}/a)$ in the different backgrounds considered.

We want to determine the CMB anisotropy spectrum induced by gravitational
waves. Using that the brightness perturbation $M$ for photons is
actually 
\[M = 4{\De T\over T} ~.\]
we obtain by integrating Eq.~(\ref{Liou.rel}) for
photons\footnote{Prior to and during recombination, photons Thomson
scatter with the electrons. For photons, we thus, in principle, have 
a collision term on  the right hand side of Eq.~(\ref{Liou.rel}). 
But this is only important on relatively small angular scales, 
$\ell\gsim 500$, which we are not
investigating here. We thus neglect the collision term.}
\be
{\Delta T \over T}(t_0, {\bf k}, {\bf n})=\exp(i{\bf k}\cdot{\bf n}t_0)
 \int_{t_{dec}}^{t_0}{n^i
n^j {\dot h}^T_{i j}(t,{\bf k})\exp(-i{\bf k}\cdot{\bf n}t)dt} ~.
\label{dT}
\ee
The power spectrum, $C_l$ of the CMB anisotropies can defined by the
expansion of the correlation function into spherical harmonics.
\be
C(\cos\theta)=\left\langle{{\Delta T} \over T}(t_0,x_0, {\bf n}) \cdot
{{\Delta T} \over T}(t_0,x_0, {\bf n}')\right\rangle_{({\bf n} \cdot {\bf n}')
=\cos\theta}=
{1 \over {4\pi}}\Sigma(2l+1)C_{\ell}P_{\ell}(\cos\theta)   \label{correl}
\ee
A somewhat lengthy calculation  relates the
gravitational wave spectrum $|\dot{h}(t,k)|^2$  via Eqs.~(\ref{dT})
and (\ref{correl}) to the $C_\ell$'s.
Taking into account the conditions (\ref{cond}), it is possible to
split this integral into
two part coming from $h^T_{11}$ and $h^T_{12}$. If we assume initially
$\langle |h^T_{11}|^2\rangle=\langle|h^T_{12}|\rangle=|H|^2$
(no polarization), the two terms are equal.
\be
C_\ell={2\over \pi}\int dkk^2|I(\ell,k)|^2
\ell(\ell-1)(\ell+1)(\ell+2) ~,
\label{Cl}\ee
with
\be
I(\ell,k)=\int_{t_{dec}}^{t_0}dt\dot{H}(t,k){j_\ell((k(t_0-t))
	\over(k(t_0-t))^2}
\label{II}
\ee
(see also \cite{AbottWise}), where $j_{\ell}$ denotes the spherical
Bessel function of order $\ell$. A self contained derivation 
of Eq.~({\ref{II}) is presented in the appendix.

Before we come to a description of the model dependence of the results, let us
discuss the expected behavior in general.
For a rough discussion we may neglect the anisotropic stresses in
Eq.~({\ref{hprime}). We first consider large scales which enter the
horizon only after decoupling, $kt_{dec}\ll 1$. These scales contribute
to $|I(\ell,k)|^2$ by roughly
$A^2k^{-3}j_{\ell}^2(kt_0)/\ell^4$. Inserting such a contribution in 
Eq.~(\ref{Cl}) and integrating over $k$ yields
\be \ell^2C_\ell  \sim A^2 ~.\ee
The integration over $k$ is only justified if the main contribution to
$j_\ell^2$ comes from the regime, where $kt_{dec}\ll 1$, in other words,
if $kt_0\sim\ell$ for a value of $k$ with 
$kt_{dec} \ll 1$, which is equivalent to $\ell \ll t_{dec}/t_0 \sim 50$.

Perturbations on small scales, $kt_{dec}\gg 1$ are damped by a factor
of about $(t_{enter}/t_{dec})^2\sim 1/(kt_{dec})^2$ until
decoupling\footnote{Here we neglect the short matter dominated period
before decoupling and approximate the damping factor by its behavior in
the radiation dominated era, $(\dot{a}/a)\sim 1/t$.}. Here
$t_{enter}\sim 1/k$ denotes the scale at which the mode $k$ enters the
horizon.
Since the main contribution to $C_\ell$ comes from the scales $k$  
with  $ kt_0 \sim \ell$, we obtain an approximate behavior of
\be \ell^2C_\ell  \sim A^2\left({50 \over\ell}\right)^4 
~~\mbox{ for } \ell \gg t_{dec}/t_0 \sim 50 ~.\ee
This analysis explains the generic behavior of the curves shown
in Fig.~3.

Let us now come to a more detailed analysis.
Having calculated the metric perturbations $h^T_{i j}$ numerically,
we can determine the CMB fluctuation spectrum according to
Eq.~(\ref{II}) by means of numerical integration (we have used 60 to
100 point Gauss-Lagrange and Gauss-Laguerre integrations) and investigate
it's dependence on the  model parameters.

\section{Results}

We have solved Eqs.~(\ref{hprime}), (\ref{Liou.rel}) and
 (\ref{Pi.rel}) numerically
for $t_{in} \leq t \leq t_0$ choosing the initial conditions
of the growing mode and  unpolarized, isotropic  waves,
 $\langle|h^T_{11}|^2\rangle=\langle|h^T_{12}|^2\rangle$.
For a given model of inflationary initial perturbations, our results
 would have to be properly weighted and added to the scalar
$C_{\ell}$'s.

We have chosen a flat initial spectrum, such that
$h^T_{in}=Ak^{-3/2}$ and ${\dot h}_{in}=0$.
\vskip 7pt

We have investigated 80 models varying the five parameters ($h_0,
\Omega_{\Lambda}, \Omega_H/\Omega_C, \beta_{\nu}, \beta_H$), where $h_0$
is the Hubble parameter $H_0$ in units of $100$km/s/Mpc,
$\beta_\nu$ denotes the number of degrees of freedom in massless
neutrini and $\beta_H$ is the corresponding number for hot dark matter
particles.

All the models lead to similar gravity wave induced
anisotropies which, for reasonable parameter choices, differ by less
than about 10\%.
The changes due to anisotropic stresses are extremely small, on the
order of 1\% or less. The main difference is caused by a non-zero
cosmological constant, which enhances the damping at late times and
thus leads to somewhat smaller perturbation amplitudes (see also
\cite{Kofman}). A similar
effect is obtained if we increase the Hubble parameter. Hot dark
matter does not induce significant changes since, at times when
the wavelengths leading to substantial CMB anisotropies enter the
horizon, hot dark matter is already non-relativistic, resulting in
nearly the same expansion law as cold dark matter. In Fig.~1, we show
the dependence of $\dot{h}(t,k)$ for fixed $k$ as a function of time
varying several  model parameters. We have chosen $k=20/t_0$, which
contributes mainly to an angle of $\theta\sim 6^o$ in the sky, or to the
$C_{\ell}$'s with $\ell \sim 10$ to $20$.
For some of these models we also
show $\dot{h}(t,k)$ as a function of $k$ for fixed time $t=t_0/2$ in Fig.~2.

\begin{figure}[htb]
\centerline{\psfig{figure=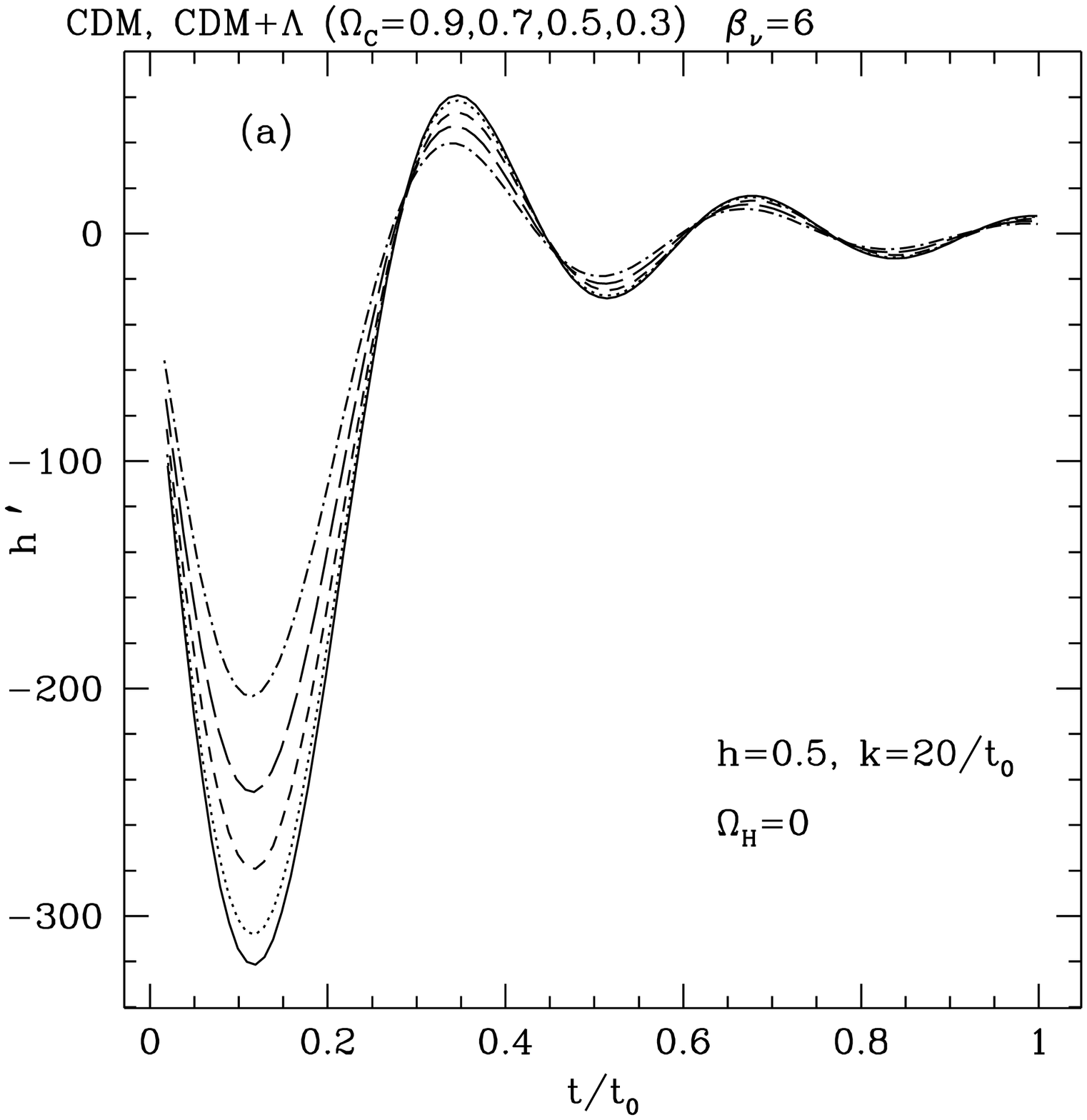,width=7.5cm}
      \psfig{figure=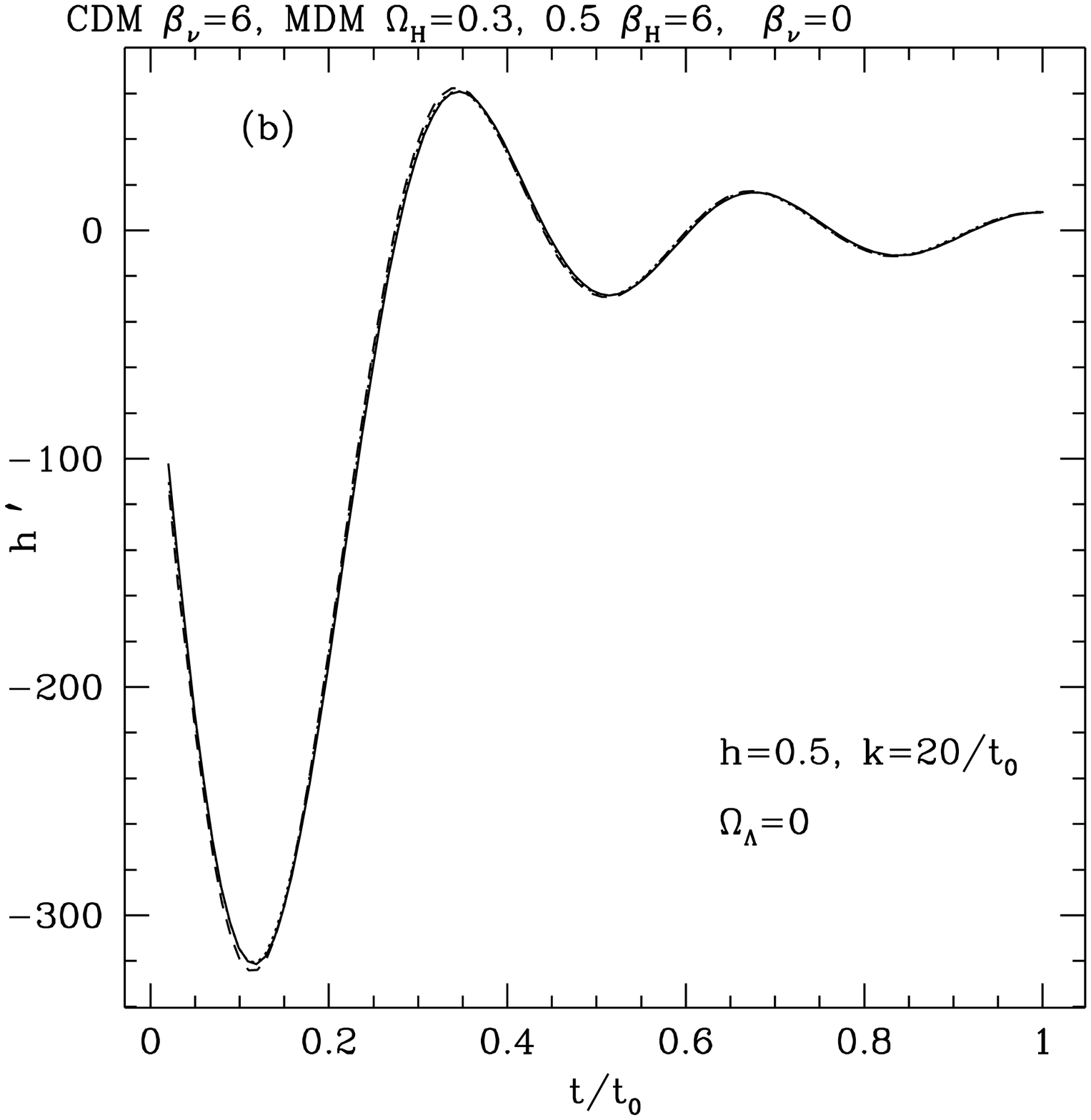,width=7.5cm}}
\centerline{\psfig{figure=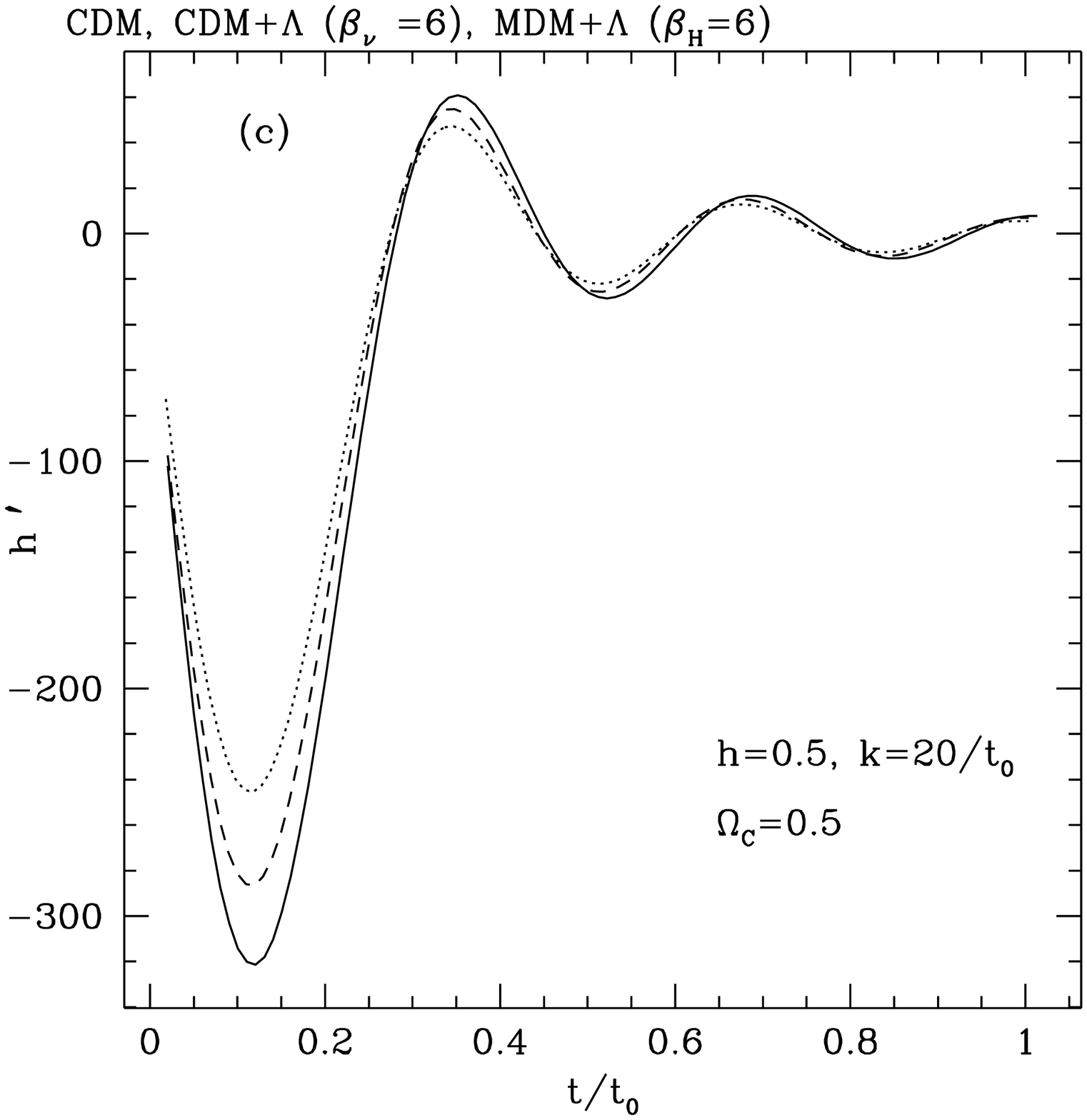,width=7.5cm}
        \psfig{figure=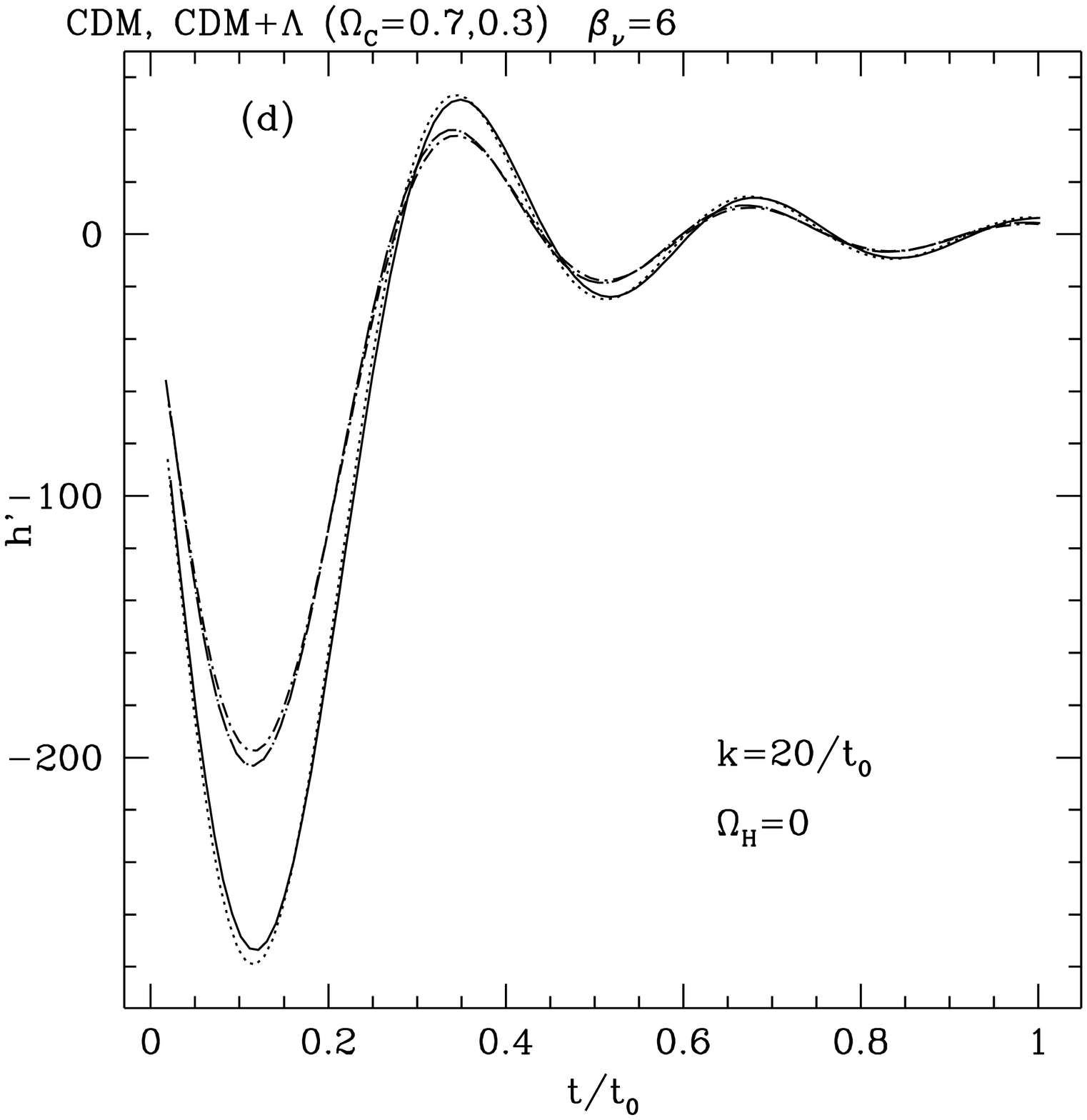,width=7.5cm}}
\caption{The variable $\dot{h}$ is shown at fixed wave number $k=20/t_0$
as function of time for
 different models. In frame (a), we consider models without HDM.
 The solid line shows pure CDM
(with 3 sorts of massless neutrini). The dotted, dashed, long dashed
and dash--dotted lines show models with increasing $\Om_\La$.
In frame (b), mixed dark matter models with $\Om_H=0.3$ (dotted) and
$\Om_H=0.5$ (dashed) are compared with standard CDM (solid line).
In frame (c) standard CDM (solid line),
with $\Om_{CDM}=\Om_\La=0.5$ (dotted line) and
$\Om_{CDM}=0.5,~ \Om_\La=\Om_H=0.25$ (dashed line) are shown. In frame
(d) we compare $\Om_\La=0.3$ models with Hubble parameters $h_0=0.5$
(dotted) and $h_0=0.75$ (solid); and  $\Om_\La=0.7$ models
with $h_0=0.5$ (dash-dotted) and $h_0=0.75$ (long-dash-dotted).}
\end{figure}

\begin{figure}[htb]
\centerline{\psfig{figure=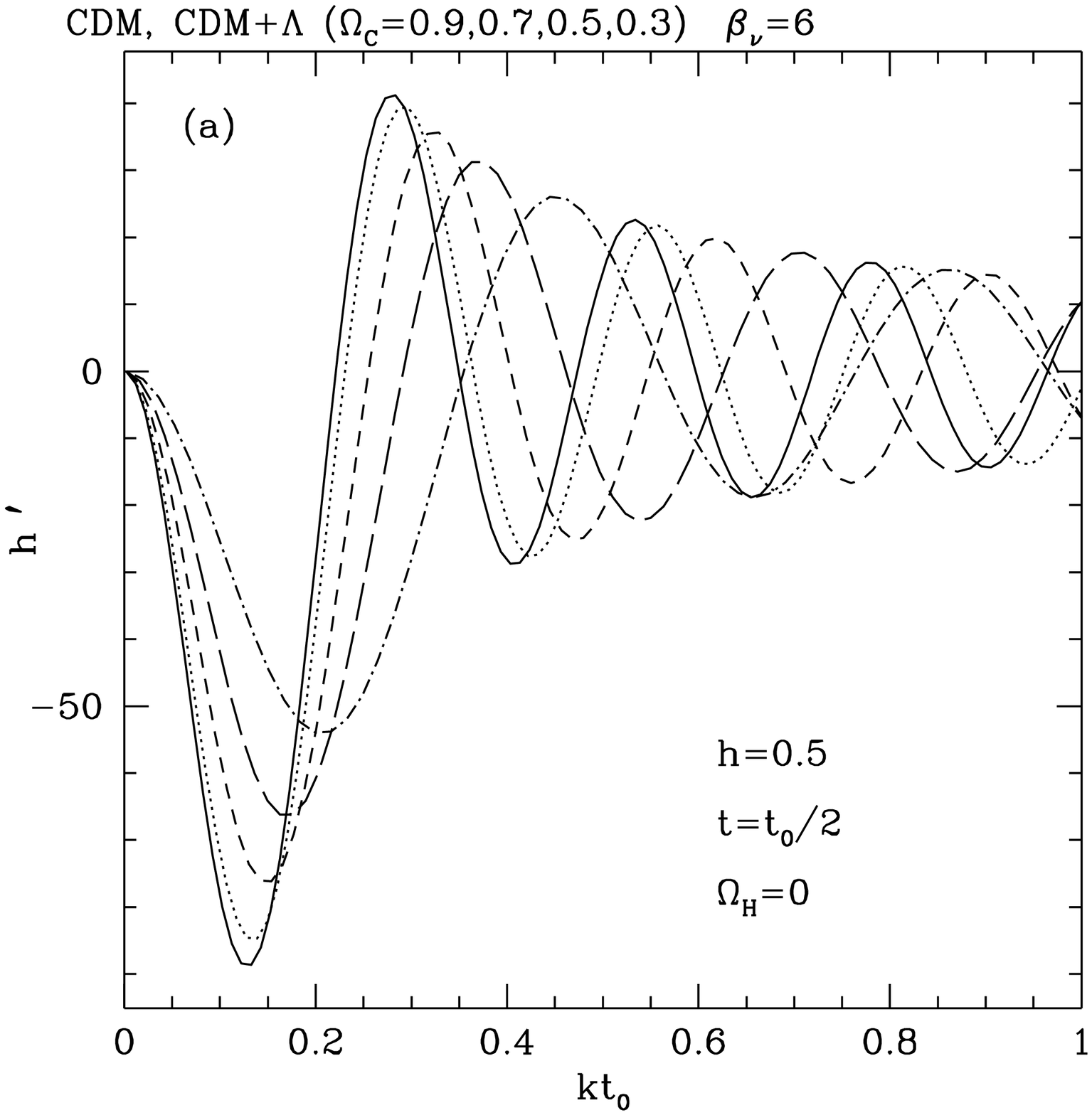,width=8cm}
      \psfig{figure=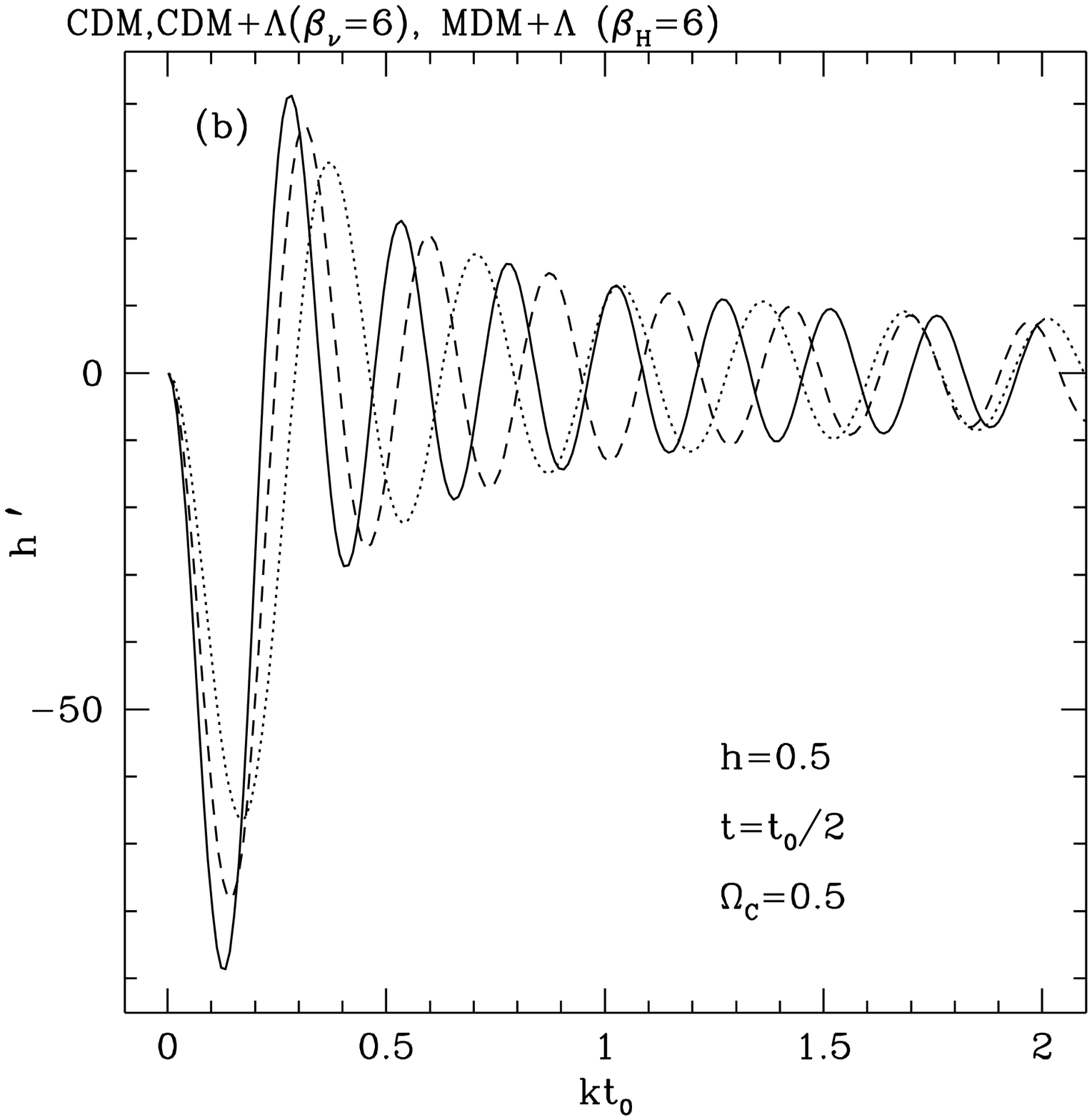,width=8cm}}
\caption{The variable $\dot{h}$ is shown at fixed time, $t=t_0/2$ as
function of the wave number $k$ for
 different models. In frame (a), we consider models without HDM.
 The solid line shows pure CDM
(with 3 sorts of massless neutrini). The dotted, dashed, long dashed
and dash--dotted lines show models with increasing $\Om_\La$.
In frame (b) standard CDM (solid line),
with $\Om_{CDM}=\Om_\La=0.5$ (dotted line) and
$\Om_{CDM}=0.5,~ \Om_\La=\Om_H=0.25$ (dashed line) are shown.}
\end{figure}

The solid line always shows
standard CDM, {\em i.e.} $\Om_C=1$ and $\beta_\nu=6$ for comparison.
The maximum amplitudes for standard CDM and CDM$+\La$ differ by more
than 30\%, while the mixed dark matter models show differences of
about 1\% only. The somewhat weaker damping due to the absence of
the anisotropic stresses provided by a massless neutrino component and
the decrease of $\dot{a}/a$
leads to the slight increase in amplitude for  mixed
dark matter models with  $\Om_H=0.5$. This result does not change
if we increase the amount of hot dark matter. However, if we decrease
$\Om_H$ to 0.3 or less no amplitude change is left and only a small
decrease in wavelengths builds up after about one oscillation. Due to
the smallness of these effects, which remain of the order of 1\% to 2\%
 when translated into the $C_{\ell}$'s, we disregard mixed dark
matter models in what follows and claim that, on the level of 1\%
accuracy, MDM and CDM lead to the same gravitational wave spectra.
Changing the number of degrees of freedom in massless neutrini or HDM
also induces very small differences of the order of 1\%.

Taking into account that an experiment always just measures the sum of
tensor and scalar contributions and first has to disentangle the
probably significantly smaller gravitational wave contribution, we
can disregard such small effects, even if the experimental error is as
small as possible, {\em i.e.} dominated by cosmic variance.

The relevant parameters to be considered are thus $\Om_M=\Om_C+\Om_H,~
\Om_\La$ and $h_0$.

In the $k$ dependence of $\dot{h}$ an additional effect comes into
play: Due to the model dependence of $t_0$, the oscillations in
$\dot{h}$ at a fixed fraction $t=ft_0$ of $t_0$ have different
wavelengths if measured in units of $t_0$. Models with a larger
 cosmological constant oscillate slower in $kt_0$ than models with small
cosmological constant. Therefore, the cancelation due to oscillations
in the integral (\ref{II}) is more severe for models with small
cosmological constant. This effect finally dominates over the larger
amplitude of $\dot{h}$ which models with large $ \La$ actually have.
In Fig.~3 we show the $C_{\ell}$ spectra for several models and in a
Fig.~4 the relative differences are indicated. The $\La$-models shown in
frames (a) of Figs.~3 and~4 show a slightly increasing amplitude with
increasing $\La$. As can be seen in frames (b) of Figs.~3 and~4, increasing
$h_0$, which does not lead to an increase in the relative oscillation
frequency, just decreases the fluctuations due to stronger
damping. The detailed model parameters of the frames (b) are just given for
information. The only parameters which really matter are $\Om_\La$ and
$h_0$. The variations induced by changing the other parameters are on
the 1\% level and thus swamped by cosmic variance.
\begin{figure}[htb]
\centerline{\psfig{figure=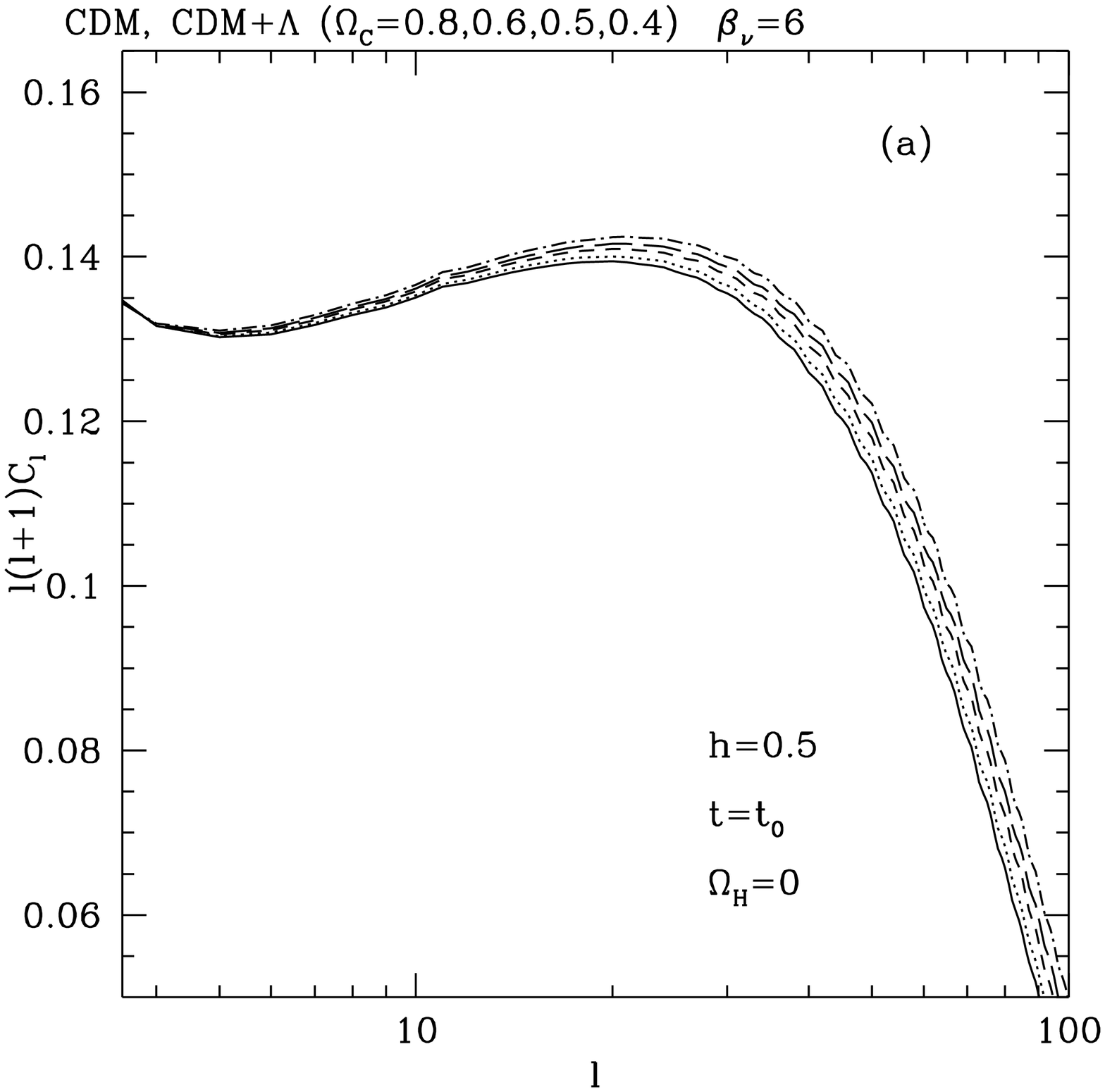,width=8cm}
      \psfig{figure=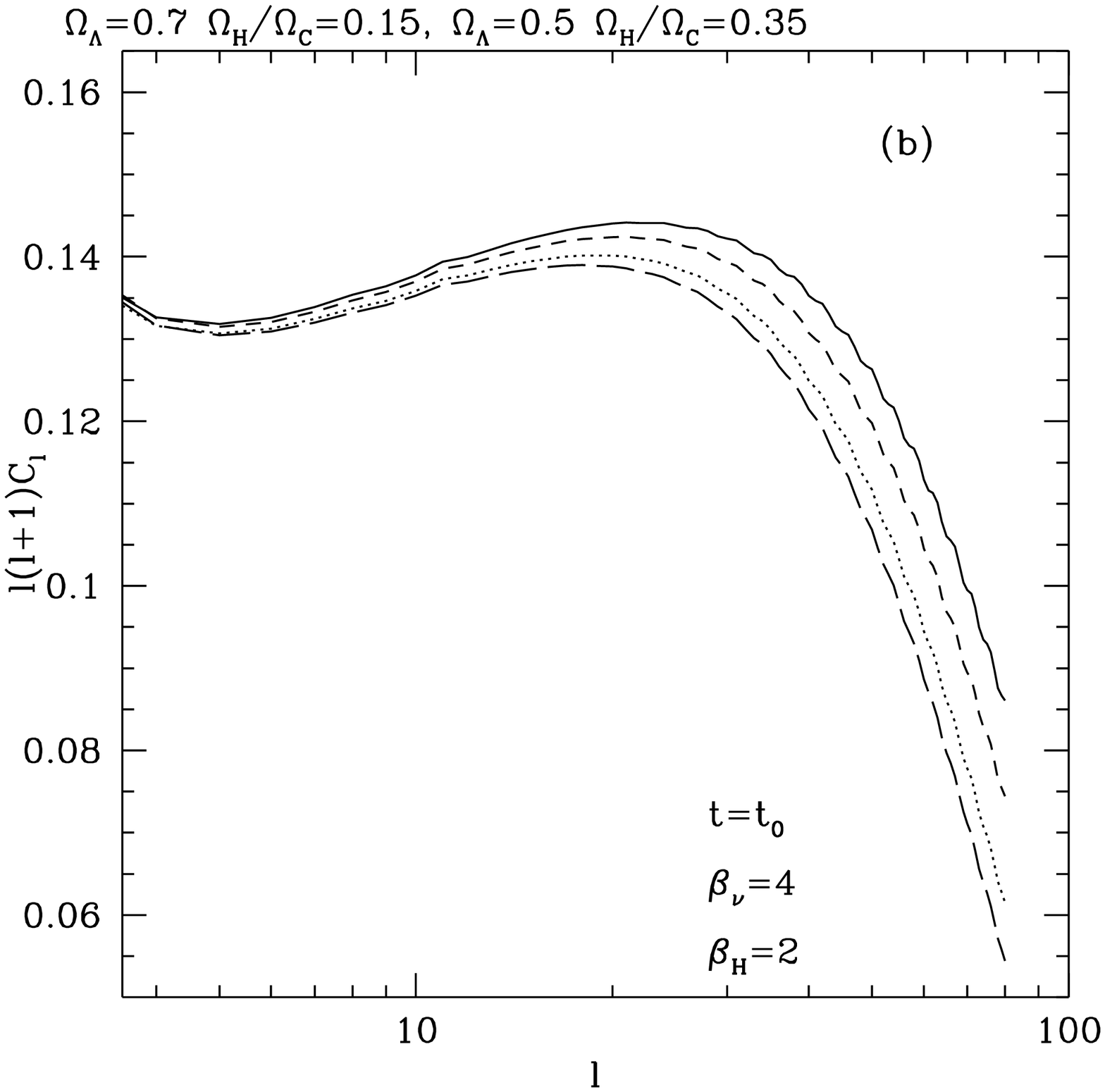,width=8cm}}
\caption{In frame (a), the angular power spectra of CMB anisotropies
induced by gravitational waves are shown
for models with different values for the cosmological constant. The
solid line represents the model  $\Om_\La=0$ (solid line) and the
amplitude increases with increasing $\Lambda$.
In frame (b), we show the effect of  increasing the Hubble
parameter. The models chosen are mixed dark matter models with
cosmological constant, $\Om_H/\Om_{CMD}=0.15,~\Om_\La=0.7$ with
$h_0=0.5$ (solid line) and $h_0=0.75$ (dotted line); and
$\Om_H/\Om_{CDM}=0.35,~ \Om_\La=0.5$ with $h_0=0.5$ (dashed line) and
$h_0=0.75$ (long dashes) are shown.}
\end{figure}
\begin{figure}[htb]
\centerline{\psfig{figure=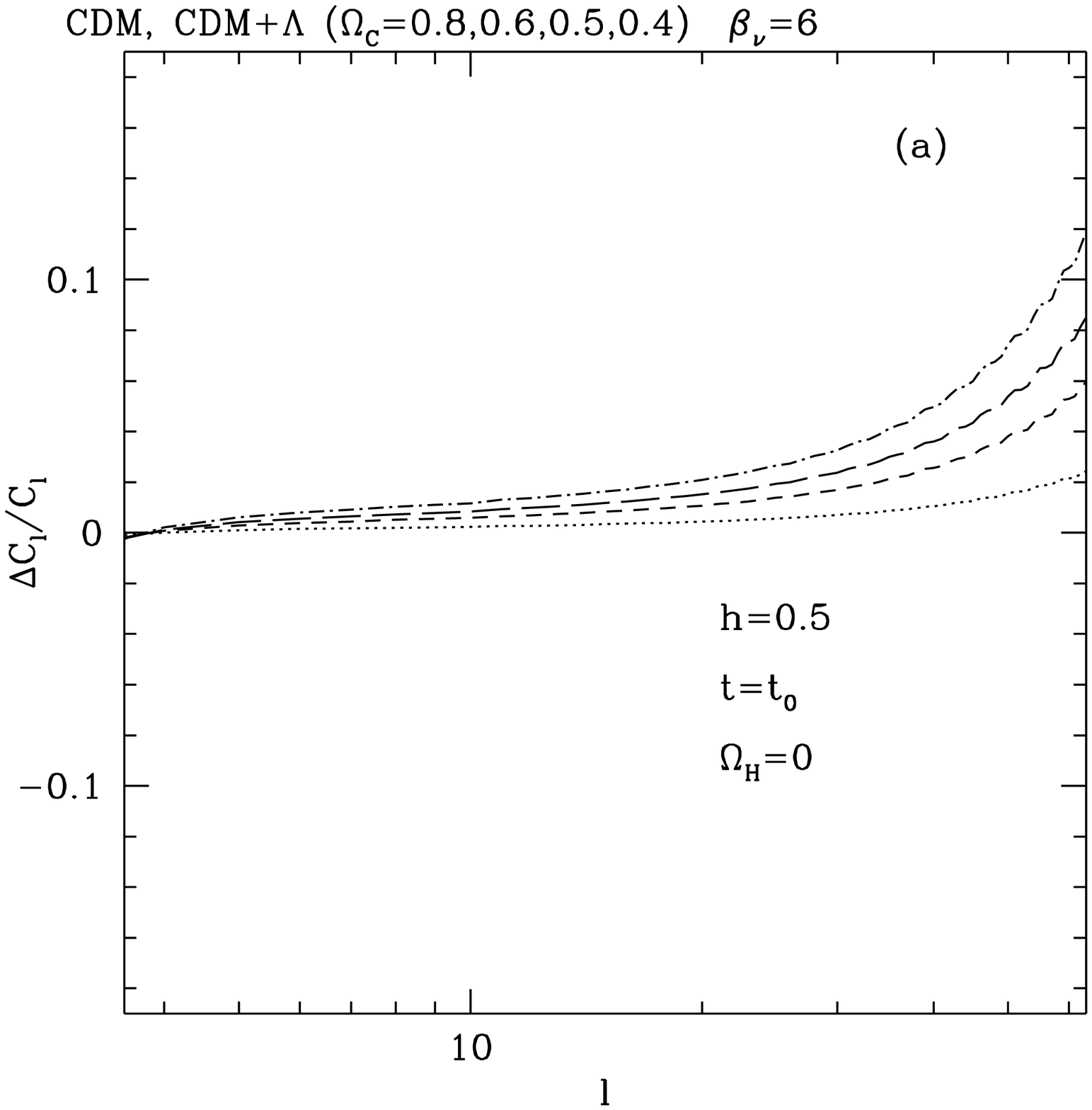,width=8cm}
      \psfig{figure=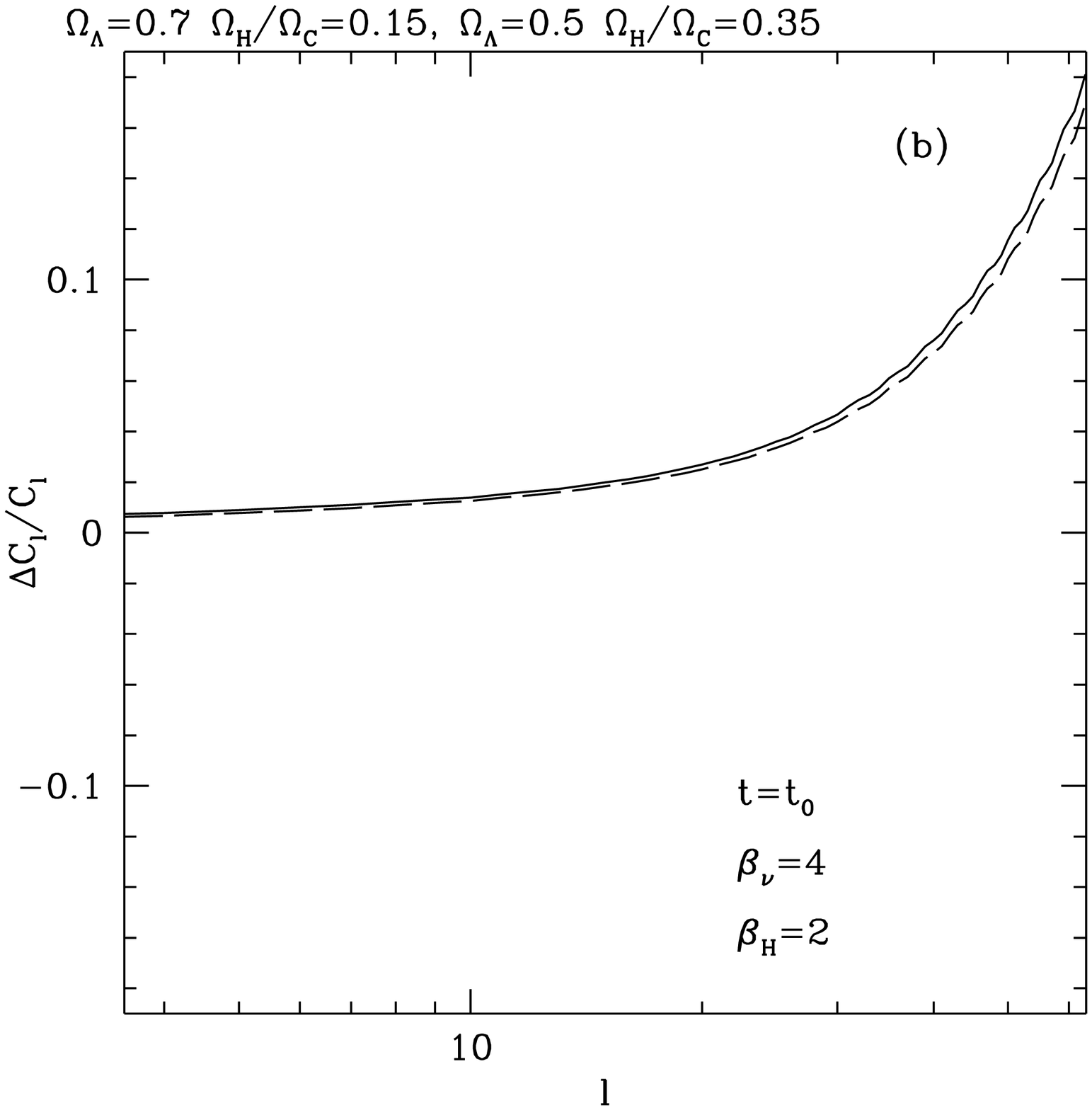,width=8cm}}
\caption{The relative differences for the models given in
Fig.~3 are shown. In frame (a) the same line types as in Fig~3a are
chosen, and the difference from standard CMB is indicated.
In frame (b) The difference between $h_0=0.5$ and $h_0=0.75$ is
shown for the model $\Om_H/\Om_{CDM}=0.35,~ \Om_\La=0.5$ (solid line)
and $\Om_H/\Om_{CMD}=0.15,~\Om_\La=0.7$ (dashed line).}
\end{figure}

The variation of the tensor $C_{\ell}$ spectrum for different
cosmological models with fixed Hubble parameter
 which are not already excluded by
other observations than CMB anisotropies never exceed 10\% for $\ell
<60$, while  variations of the Hubble parameter can lead to changes in the
spectrum of up to 15\%.

\section{Conclusions}
We have calculated the tensor contribution to the CMB anisotropies in mixed
dark matter models with and without cosmological constant. We have
included a previously neglected source term in the evolution equation
for metric perturbations. Our findings are however quite modest:
By reasons of cosmic variance, the statistical relative error in
$C_{\ell}$ measured from only one point in the universe is always
$1/\sqrt{2\ell+1}$. This is a very significant uncertainty, especially
for the gravitational wave contribution which peaks around $\ell\sim
20$ and has already dropped by a factor of about 2 at $\ell=60$
(see Fig.~3).

In non
of the considered models the influence of the anisotropic stress
source becomes large enough to induce a difference in the $C_{\ell}$
spectrum which is larger than cosmic variance. The same is true for
hot dark matter contributions. Only an extremely large cosmological
constant or a difference in the Hubble parameter can induce changes in
the gravitational wave spectrum which are in principle observable but
nevertheless small.

This finding has one negative and one positive aspect: Unfortunately, the
gravitational wave contribution does not contain detailed information
about the cosmological parameters considered here and can thus not
be used to measure them with high accuracy. On the other hand, since
this contribution is so model independent, it conserves its
information about the initial condition and thus about the amplitude
and spectral index which it inherited during, e.g., an inflationary epoch.
\vspace{0.5cm}

\noindent{\bf Acknowledgments:}\hspace{0.2cm}
T. Kanhiashvili would like to
express her thanks to the University of Geneva for hospitality. T.K.
is grateful to R. Valdarnini and H. Miheeva for helpful remarks.
It is a pleasure to thank also A. Melchiorri and N. Straumann
for useful discussions. This
work was partially supported by the Swiss National Science Foundation.
\vspace{1.3cm}
\newpage
\setcounter{equation}{0}
\renewcommand{\theequation}{A\arabic{equation}}
\appendix
{\Large \bf Appendix: The $C_\ell$'s from gravitational waves}
\vspace{0.2cm}\\
We consider metric perturbations which are produced by some isotropic
random process (for example during inflation). After production, they
evolve according to a deterministic equation of motion. 
The correlation functions of $h_{ij}({\bf k},t)$  have to be of the form
\bea
\langle h_{ij}({\bf k},t) h^*_{lm}({\bf k},t')\rangle &=& 
  [k_ik_jk_lk_mH_1(k,t,t') + \nonumber\\  &&
  (k_ik_l\de_{jm}+k_ik_m\de_{jl}+k_jk_l\de_{im}
 +k_jk_m\de_{il})H_2(k,t,t') + \nonumber \\ &&
k_ik_j\de_{lm}H_3(k,t,t') +k_lk_m\de_{ij}H_3^*(k,t',t) +
\nonumber\\ &&
+\de_{ij}\de_{lm}H_4(k,t,t') +
(\de_{il}\de_{jm}+\de_{im}\de_{jl})H_5(k,t,t')] ~.
\label{hijlmansatz}
\eea
Here the functions $H_a$ are functions of the modulus $k=|{\bf k}|$ only.
Furthermore, all of them except $H_3$ are hermitian in $t$ and $t'$.
This is the most general ansatz for an isotropic correlation tensor
satisfying the symmetries required. To project out the tensorial part
of this correlation tensor we act on $h_{ij}$ it with the tensor projection
operator,
\bea
 T_{ij}^{~~ab}&=&(P_{il}P_{jm}-(1/2)P_{ij}P_{lm})P^{ma}P^{lb}~~\mbox{ with}\\
P_{ij} &= & \de_{ij}-\hat{k}_i\hat{k}_j~.
\eea
This yields
\bea
\lefteqn{\langle h^{(T)}_{ij}({\bf k},t)h^{(T)*}_{lm}({\bf
	k},t')\rangle =} \nonumber \\
&&	H_5(k,t,t')[\de_{il}\de_{jm}+\de_{im}\de_{jl}   -\de_{ij}\de_{lm} + 
	k^{-2}(\de_{ij}k_lk_m + \nonumber \\ 
&&	\de_{lm}k_ik_j -\de_{il}k_jk_m - \de_{im}k_lk_j -\de_{jl}k_ik_m
	-\de_{jm}k_lk_i) + \nonumber \\
&&	k^{-4}k_ik_jk_lk_m] \label{Ctau}
    ~.\eea
From Eq.~(\ref{dT}), we then obtain
\bea
&& \left\langle{\Delta T \over T}( {\bf n}){\Delta T \over T}( {\bf
n}')\right\rangle \equiv
 {1\over V}\int d^3x \left({\Delta T \over T}({\bf n,x})
	{\Delta T \over T}( {\bf n',x})\right) =
 \nonumber \\  && \left({1\over 2\pi}\right)^3
\int k^2dkd\Om_{\hat{\bf k}}\int_{t_{dec}}^{t_0}dt  \int_{t_{dec}}^{t_0}dt'
\exp(i{\bf k}\cdot{\bf n}(t_0-t))
\exp(-i{\bf k}\cdot{\bf n}(t_0-t')) \cd \nonumber\\ &&
\left[\langle{\dot h}^{(T)}_{i j}(t,{\bf k}){\dot h}_{lm }^{(T)*}(t',{\bf k})
	\rangle n_in_jn'_ln'_m \right] ~.   \label{dT2}
\eea
Here $d\Om_{\hat{\bf k}}$ denotes the integral over directions in $\bf
k$ space. We use the normalization of the Fourier
transform 
\[ \hat f({\bf k})={1\over \sqrt{V}}\int d^3x \exp(i{\bf x\cd k})f({\bf x}) ~,~~~~
  f({\bf x})={1\over (2\pi)^3}\int d^3k \exp(-i{\bf x\cd k})
	\hat f({\bf k}) ~,\]
where $V$ is an (arbitrary) normalization volume.

We now introduce the form (\ref{Ctau}) of $<h^{(T)}h^{(T)}>$. 
We further make use
of the assumption that the perturbations have been created at some
early epoch, e.g. during an inflationary phase, after which they
evolved deterministically. The function $H_5(k,t,t')$ is thus a
product of the form
\be  H_5(k,t,t') = H(k,t)\cd H^*(k,t') ~.\ee
Introducing this in Eq.~(\ref{dT2})  yields
\bea
\lefteqn{ \left\langle{\Delta T \over T}( {\bf n}){\Delta T \over T}
	( {\bf n}')\right\rangle=}
 \nonumber \\ && \left({1\over 2\pi}\right)^3
\int k^2dkd\Om_{\hat{\bf k}}
\left[({\bf n}\cd{\bf n}')^2 -1+\mu'^2+\mu^2-
	4\mu\mu'({\bf n}\cd{\bf n}')+\mu^2\mu'^2\right] \cd \nonumber\\
&&	\int_{t_{dec}}^{t_0}dt  
\int_{t_{dec}}^{t_0}dt' \left[\dot H(k,t)\dot H^*(k,t')
\exp(ik\mu(t_0-t))\exp(-ik\mu'(t_0-t')) \right]~, \label{dT2.5}
\eea
where $\mu=(n\cd\hat{\bf{k}})$ and  $\mu'=(n'\cd\hat{\bf{k}})$.
To proceed, we  use the identity \cite{AbSt}
\be
\exp((ik\mu(t_0-t))=\sum_{r=0}^{\infty}(2r+1)i^rj_r(k(t_0-t))P_r(\mu)~.
\ee
Here $j_r$ denotes the spherical Bessel function of order $r$ and
$P_r$ is the Legendre polynomial of degree $r$.

Furthermore, we replace each factor of $\mu$ in Eq.~(\ref{dT2.5})
by a derivative of the
exponential $\exp(ik\mu(t_0-t))$ with respect to $k(t_0-t)$ and
correspondingly with $\mu'$. We then obtain
\bea
\lefteqn{\left\langle{\Delta T \over T}( {\bf n}){\Delta T \over T}
	( {\bf n}')\right\rangle=}
 \nonumber \\ && \left(1\over 2\pi\right)^3 
\sum_{r,r'}(2r+1)(2r'+1)i^{(r-r')}\int k^2dkd\Om_{\hat{\bf k}}
P_r(\mu)P_{r'}(\mu') \times
 \nonumber \\ && 
\Big[2 ({\bf n}\cd{\bf n}')^2 
      \int dtdt'j_r(k(t_0-t))j_{r'}(k(t_0-t'))\dot H(k,t)\dot H^*(k,t')
 \nonumber \\ && 
-\int dtdt'[j_r(k(t_0-t))j_{r'}(k(t_0-t'))+
	j_r''(k(t_0-t))j_{r'}(k(t_0-t')) +
 \nonumber \\ && 
	j_r(k(t_0-t))j_{r'}''(k(t_0-t'))-  
	j_r''(k(t_0-t))j_{r'}''(k(t_0-t'))
]\dot H(k,t)\dot H^*(k,t')
 \nonumber \\ && 
-4({\bf n}\cd{\bf n}') \int dtdt'
	j_r'(k(t_0-t))j_{r'}'(k(t_0-t'))
	\dot H(k,t)\dot H^*(k,t')\Big]~. \label{dT3}
\eea
Here only the Legendre polynomials, $P_r(\mu)$ and  $P_{r'}(\mu')$  depend
 on the direction $\hat{\bf k}$. To perform the integration
$d\Om_{\hat{\bf k}}$, we use the addition theorem for the spherical
harmonics $Y_{rs}$,
\be
P_r(\mu)={4\pi\over (2r+1)}   \label{add}
	\sum_{s=-r}^rY_{rs}({\bf n})Y^*_{rs}(\hat{\bf k})~.
\ee
 The orthogonality of the spherical harmonics then yields
\bea
\lefteqn{(2r+1)(2r'+1)\int d\Om_{\hat{\bf k}}P_r(\mu)P_{r'}(\mu')=}
\nonumber \\ &&
 16\pi^2\de_{rr'}\sum_{s=-r}^rY_{rs}({\bf n})Y^*_{rs}({\bf n}') =
\nonumber \\ &&
 4\pi\de_{rr'}P_r({\bf n}\cd{\bf n}') ~. \label{Yorth}
\eea
In Eq.~(\ref{dT3}) the integration over $d\Om_{\hat{\bf k}}$ then
leads to  terms of the form
$({\bf n}\cd{\bf n}')P_r({\bf n}\cd{\bf n}')$ and 
$({\bf n}\cd{\bf n}')^2P_r({\bf n}\cd{\bf n}')$. To reduce them, we use
\be
xP_r(x)={r+1\over 2r+1}P_{r+1} +{r\over 2r+1}P_{r-1}~.
\ee
Applying this and its iteration for $x^2 P_r(x)$, we obtain

\bea
\lefteqn{\langle{\Delta T \over T}( {\bf n}){\Delta T \over T}^* ({\bf
	n}') \rangle=}
 \nonumber \\ && {1\over 2\pi^2} 
 \sum_{r}(2r+1)\int k^2dk \int dtdt' \dot H(k,t)\dot H^*(k,t')\Big\{
 \nonumber \\ &&  \left[ 
{2(r+1)(r+2)\over (2r+1)(2r+3)} P_{r+2}+{1 \over (2r-1)(2r+3)}P_{r}
+ {2r(r-1)\over (2r-1)(2r+1)}P_{r-2} \right]\times
\nonumber \\ &&
       j_r(k(t_0-t))j_{r}(k(t_0-t')) - P_r [j_r(k(t_0-t)j_r''(k(t_0-t'))
\nonumber \\ &&  + j_r(k(t_0-t'))j_r''(k(t_0-t))
                 -j_r''(k(t_0-t))j_{r'}''(k(t_0-t'))]
 \nonumber \\ && 
-4 \left[{r+1 \over 2r+1}P_{r+1} + {r \over 2r+1}P_{r-1}\right]
	j_r'(k(t_0-t))j_{r}'(k(t_0-t'))
 \Big\}~, \label{dT4}
\eea
where the argument of the Legendre polynomials, $\bf n\cd n'$, has
been suppressed.
Using the relations 
\be
j_r'=-{r+1\over 2r+1}j_{r+1} +{r\over 2r+1}j_{r-1}
\ee
for Bessel functions,
and its iteration for $j''$, we can rewrite Eq.~(\ref{dT4}) in terms of
the Bessel functions $j_{r-2}$ to $j_{r+2}$.

To proceed we use the definition of $C_\ell$:
\be
\left\langle{{\Delta T} \over T}({\bf n}) \cdot
{{\Delta T} \over T}({\bf n}')\right\rangle_{({\bf n} \cdot {\bf n}')
=\cos\theta}=
{1 \over {4\pi}}\Sigma(2l+1)C_{\ell}P_{\ell}(\cos\theta)   \label{correlAp}
\ee
If we expand 
\be 
{\Delta T \over T}({\bf n})=\sum_{\ell,m}a_{\ell,m}Y_{\ell,m}({\bf n})
\label{expan}
\ee
and use the orthogonality of the spherical harmonics as well as the
addition theorem, Eq.~(\ref{add}), we get
\be 
   C_\ell = \langle a_{\ell,m}a^*_{\ell,m}\rangle ~. \label{Cell}
\ee
We thus have to determine the correlators
\be
\langle a_{\ell m} a^*_{\ell' m'} \rangle =
\int d\Om_{\bf n} \int d\Om_{\bf n'}
\left\langle {\Delta T \over T}^* {\Delta T \over T} \right\rangle  
Y^*_{\ell m} ({\bf n}) Y_{\ell 'm'} ({\bf n '}) ~.  \label{all'}
\ee 
Inserting our result~(\ref{dT4}), we obtain  the somewhat lengthy expression
\bea \lefteqn{\langle a_{\ell m} a^*_{\ell' m'} \rangle =}
\nonumber \\ &&
{2\over \pi} \delta_{\ell \ell'} \delta{mm'} \int dk k^2 
\int dt dt' \dot H(k,t)\dot H^*(k,t')\Big\{
\nonumber \\ &&    j_l(k(t_0-t))j_l(k(t_0-t')) \times  
\nonumber \\ &&
      \left (
        {1 \over (2 \ell-1)(2 \ell +3)} + {2 (2 \ell^2+2 \ell -1) \over 
        (2 \ell-1)(2 \ell +3)} + 
        {(2 \ell^2+2 \ell -1)^2 \over (2 \ell-1)^2(2 \ell +3)^2} \right. 
\nonumber \\ && \left.
       -{4 \ell^3 \over (2 \ell-1)^2(2 \ell +1)} -
       {4 (\ell+1)^3 \over (2 \ell+1)(2 \ell +3)^2}  \right ) 
\nonumber \\ && 
-\left[j_{\ell}(k(t_0-t))j_{\ell+2}(k(t_0-t'))
+j_{\ell+2}(k(t_0-t))j_{\ell}(k(t_0-t'))\right] \times 
\nonumber \\ && {1\over 2l+1}
     \left ( {2(\ell+2)(\ell+1)(2\ell^2+2\ell-1) 
              \over (2\ell-1)(2\ell+3)^2}
            +{2(\ell+1)(\ell+2) \over (2\ell+3)}
     -{8(\ell+1)^2(\ell+2) \over (2\ell+3)^2} \right )
\nonumber \\ && 
-\left[j_{\ell}(k(t_0-t))j_{\ell-2}(k(t_0-t'))
+j_{\ell-2}(k(t_0-t))j_{\ell}(k(t_0-t'))\right] \times 
\nonumber \\ &&{1\over 2l+1}
\left ( {2 \ell (\ell-1)(2\ell^2+2\ell-1) \over (2\ell-1)^2(2\ell+3)}
+{2 \ell (\ell-1) \over (2\ell-1)(2}
 -{8 \ell^2 (\ell-1) \over (2\ell-1)^2} \right )
\nonumber \\ &&
+j_{\ell+2}(k(t_0-t))j_{\ell+2}(k(t_0-t')) \times
\nonumber \\ && 
\left ( {2(\ell+2)(\ell+1) \over (2\ell+1)(2\ell+3)}-
{4(\ell+1)(\ell+2)^2 \over (2\ell+1)(2\ell+3)^2} + 
{(\ell+1)^2(\ell+2)^2 \over (2\ell+1)^2(2\ell+3)^2} \right )
\nonumber \\ && 
+j_{\ell-2}(k(t_0-t))j_{\ell-2}(k(t_0-t')) \times
\nonumber \\ &&
\left. \left ( {2 \ell (\ell-1) \over (2\ell-1)(2\ell+1)}-
{4 \ell (\ell-1)^2 \over (2\ell-1)^2 (2\ell+1)} + 
{\ell^2(\ell-1)^2 \over (2\ell-1)^2(2\ell+1)^2} \right )
\right \}~\label{dT5}
\eea

An analysis of the coefficient of each term reveals that this
result is equivalent to Eq.~(\ref{Cl}) with.
\bea
I(\ell,k) &=&     {j_{\ell+2}(k(t_0-t))\over (2\ell+3)(2\ell+1)}
        + {2j_{\ell}(k(t_0-t))\over (2\ell+3)(2\ell-1)}
        +{j_{\ell-2}(k(t_0-t))\over (2\ell+1)(2\ell-1)}y\\
        &=& {j_{\ell}(k(t_0-t))\over (k(t_0-t))^2}~.
\eea

\end{document}